%% file: pyeb_abz.tex
\documentclass[10pt]{llncs} 

\usepackage{hyperref}
\usepackage{color}
\usepackage{listings}
\usepackage{amsfonts}
\usepackage{xspace}
\usepackage{flushend}
\usepackage {bsymb}
\usepackage[T1]{fontenc}

\definecolor{darkgreen}{rgb}{0,0.6,0}
\definecolor{darkblue}{rgb}{0,0,0.55}
\definecolor{darkred}{rgb}{0.4,0.0,0}
\definecolor{darkgray}{rgb}{0.66,0.66,0.66}
\definecolor{darkpink}{rgb}{0.91, 0.33, 0.5}
\definecolor{lightgray}{rgb}{0.95,0.95,0.92}
\definecolor{darkpurple}{rgb}{0.59, 0.44, 0.84}

\definecolor{pred}{rgb}{0.6,0,0}
\definecolor{pgrey}{rgb}{0.75,0.75,0.75}

\newcommand{\pyeb}{\texttt{pyeb}\xspace}
\newcommand{\mypy}{\texttt{mypy}\xspace}
\newcommand{\pylint}{\texttt{py\-lint}\xspace}
\newcommand{\pyflakes}{\texttt{py\-flakes}\xspace}
\newcommand{\prospector}{\texttt{Pros\-pec\-tor}\xspace}
\newcommand{\bandit}{\texttt{Ban\-dit}\xspace}
\newcommand{\nagini}{\texttt{Na\-gi\-ni}\xspace}

\newcommand{\viper}{\texttt{Vi\-per}}
\newcommand{\pymc}{\texttt{Py\-Model\-Check\-ing}\xspace}
\newcommand{\pip}{\texttt{pip}\xspace}
\newcommand{\pypi}{\texttt{PyPI}\xspace}

\newcommand{\eb}{Event-B\xspace}

\newcommand{\ebkeyw}[1]{\textsf{\bf #1}}
\newcommand{\ebtag}[1]{\textcolor{pgrey}{\textsc{\bf #1}}}
\newcommand{\ebcode}[1]{\textsf{#1}}

\newcommand{\lcb}{{\tt {\char '173}}}
\newcommand{\rcb}{{\tt {\char '175}}}

\begin{document}
\title{\pyeb: A Python Implementation of Event-B Refinement Calculus}

\author{N{\'e}stor Cata{\~n}o}

\institute{ 
Faculdade de Ciências e Tecnologia \\
Universidade do Algarve\\ Portugal \\
\email{nestor.catano@gmail.com}
}

\maketitle
\thispagestyle{empty}

\input{abstract}
\input{intro}

\input{prelim}
\input{encoding}

\input{exp}
\input{related}
\input{conclusion}

\bibliographystyle{plain}
\bibliography{pyeb_abz}

\end{document}

%% file: abstract.tex
\begin{abstract}
  
  This paper presents the \pyeb tool, a Python implementation of the
  \eb refinement calculus. The \pyeb tool takes a Python program and
  generates several proof obligations that are then passed into the Z3
  solver for verification purposes. The Python program represents an \eb model. Examples of these proof
  obligations are machine invariant preservation, feasibility of
  non-deterministic event actions, event guard strengthening, event
  simulation, and correctness of machine variants. The Python program
  follows a particular object-oriented syntax; for example, actions,
  events, contexts, and machines are encoded as Python classes. We implemented \pyeb as a \pypi (Python Package Index) library, which is freely available online. We carried out a case study on the use of \pyeb. We modelled and verified several sequential
  algorithms in Python, e.g., the binary search algorithm and the square-root algorithm, among others. Our experimental results show that \pyeb
  verified the refinement calculus models written in Python.
  
\end{abstract}

%% file: intro.tex
\section{Introduction}
\label{sec_intro}

Formal methods are a set of mathematically-based techniques that
ensure the reliability of software systems. The adoption of formal
methods in the software industry remains low. This problem is
exacerbated by the lack of mathematical background in formal methods
techniques and tools of software engineers. In particular, the use of
refinement techniques with \eb~\cite{abrial:book:10} requires
specialised knowledge of predicate and refinement calculus. Our work
seeks to help software developers address this issue while writing
program and system specifications in a popular mainstream program
language such as Python.

This paper presents \pyeb, a Python implementation of \eb refinement
calculus in which users can write and verify \eb models written in
object-oriented Python. The \pyeb tool is implemented as a \pypi
(Python Package Index) library~\cite{pypi}. \pyeb takes a Python
program and generates various proof obligations similar to the ones
generated by the Rodin IDE for \eb models~\cite{abrial:book:10}. \pyeb
uses the Z3 STM solver~\cite{z3py} to discharge the generated proof
obligations. Their correctness attests to the correctness of the
Python models. Examples of these proof obligations are machine
invariant preservation, feasibility of non-deterministic event
actions, event guard strengthening, event simulation, and correctness
of machine variants.

\medskip

The \textbf{main contributions} of this paper are the following:

\begin{itemize}
\item We have carried out an implementation of the refinement calculus in Python. Our implementation (the \pyeb tool) is a \pypi
  library~\cite{pypi}, which can be installed using the \pip~\cite{pip} Python package installer. \pyeb supports \eb's syntax, including deterministic and non-deterministic event
  actions, events, machines, machine contexts, machine variants and
  invariants, events, and, event and machine refinement.
\item We have performed a case study on the modelling and verification
  of sequential algorithms with \pyeb. The algorithms have been
  written elsewhere by J.-R.~Abrial in~\cite{abrialseq:09}. They
  include the binary search algorithm, finding the minimum element of
  an array, searching for a value in an array, and calculating the
  square root.
\item \pyeb contributes to our long-term goal of making
  refinement-calculus-based formal methods more accessible and popular
  for standard programmers unfamiliar with \eb's mathematical
  syntax. 
\end{itemize}

\textbf{Organisation.}  This paper is organised as follows:
Section~\ref{sec_prel} introduces the \eb language, its syntax and
semantics. Section~\ref{sec_encoding} discusses our approach to the modelling
\eb in Python. Section~\ref{sec_exp} discusses the challenges of the
case study. Section~\ref{sec_rel} discusses related work. Finally,
Section~\ref{sec_conc} concludes and discusses future work.

%% file: prelim.tex
\section{Preliminaries}
\label{sec_prel}

\subsection{\eb}
\label{sec_eb} \eb\ models represent the development of transition
systems. The models comprise \emph{machines} and
\emph{contexts}. Machines comprise a static part that defines
observations about the system, and a series of state transition
operations called \emph{events}. Each operation must maintain the
machine invariants. Contexts define the static part of a machine
(uninterpreted sets, constants, axioms, and theorems), whereas
variables and invariants are defined within the machines. A
machine definition generates a series of proof obligations,
basically, theorems stating a property that must be true for the
machine to be consistent. For example, if an \eb model includes an
invariant property that states that some mathematical relation $r$
is a function, then for each machine event that modifies $r$, a
proof obligation is generated that states that the modified
relation $r'$ must be a function.

Software development with \eb relies on \emph{model
refinement}~\cite{Abrial:Ref:07,roever:dref:98} whereby a machine goes
through a series of stages, each adding more details (called
observations) to the description of the system. Hence, the most
abstract machine is first developed and verified to satisfy the
system's safety properties. Then, with every refinement machine, one
must prove a correspondence between the refinement and its abstract
machine, therefore, the refinement (concrete) machine fulfils at least
the same properties that the abstract machine does. To prove such
correspondence, refinement proof obligations are to be discharged
(proven) to ensure that each refinement is a faithful and sound model of
the previous machine, and so that all the machines satisfy the safety
properties of the most abstract machine.


Figure \ref{fig_evt} shows the syntax for events and event
refinements. An event can have one of three possible $status$:
\ebkeyw{ordinary}, \ebkeyw{convergent}, or
\ebkeyw{anticipated}. \ebkeyw{convergent} events and machine variants
are used to model systems that halt, e.g., the sequential algorithms
presented in Section~\ref{sec_exp}. The body of an event (written
between the keywords \ebkeyw{then} and \ebkeyw{end}) is composed of a
set of actions. Actions compute new values for machine variables, thus
performing an observable state transition. Actions can either be
deterministic or non-deterministic and may change the value of
machine variables as part of state transitions. 
\eb\ provides some simplified syntax versions. For example, if
$x$ is empty, the event does not include an \ebkeyw{any} clause. The event does not include a clause \ebkeyw{then} if the actions are missing. 

\begin{figure}[t]
\begin{center}
\begin{tabular}{c@{\hspace*{12pt}}c}
\begin{tabular}{l} 
$status$~$evt_0$ \\
~\ebkeyw{any} $x$\\
~\ebkeyw{where} $G(s,c,v,x)$ \\
~\ebkeyw{then} \\
~~~$v :\!|\;BA_0(s,c,v,x,v')$ \\
\ebkeyw{end} 
\end{tabular} &
\begin{tabular}{l} 
$status$~$evt$ \\
\ebkeyw{refines}~$evt_0$ \\
~\ebkeyw{any} $y$\\
~\ebkeyw{where} $H(s,c,w,y)$ \\
~\ebkeyw{with} $a:P(a)$\\
~\ebkeyw{then} \\
~~$w :\!|\;BA(s,c,w,y,w')$ \\
\ebkeyw{end}
\end{tabular}
\end{tabular}
\end{center}
\caption{$i.)$ event syntax $ii.)$ event refinement syntax}
\label{fig_evt}
\end{figure}

%% file: encoding.tex
\section{Python Encoding}
\label{sec_encoding}

Next, we discuss our Python implementation of the refinement
calculus. We use Z3's Python API~\cite{z3py} to represent
uninterpreted constants and functions. For example, for the binary
search algorithm discussed in Section~\ref{sec_binary}, we use an
integer-valued function \texttt{f}, an integer constant \texttt{n}
(representing the size of \texttt{f}) and the value \texttt{v} we
search in \texttt{f}. 

\smallskip
\begin{lstlisting}[language=Python,morekeywords={assert,self}]
self.f = Function('f',IntSort(),IntSort()) 
self.n = Const('n',IntSort()) 
self.v = Const('v',IntSort())
\end{lstlisting}
\smallskip

We use Python object-oriented syntax to encode the \eb syntax. Class
\texttt{B\-Context} (not shown here) encodes machine contexts. This
class encodes constants, axioms, and theorems as class fields of type
\emph{dictionary}. \texttt{B\-Context} uses setter and getter methods to add
and retrieve any value from or to these dictionaries. \eb constants
and functions are encoded as Z3 constants and functions of some
particular sort, e.g. \texttt{IntSort()} above.

\medskip

\noindent \textbf{Primed variables.} \eb models represent the
development of discrete transition systems. We encode $x'$ (the next
value of $x$) in Python with the help of the function \texttt{prime}
below.

\smallskip
\begin{lstlisting}[language=Python,morekeywords={assert,self}]  
def prime(x):
 s = x.sort()
return Function("prime", s, s)(x)
\end{lstlisting}
\smallskip

Figure~\ref{fig_inc} presents an excerpt of the \eb model of the
binary search algorithm. The left side shows the abstract event
\ebcode{progress} together with the invariants of the abstract
machine. The right side shows its refinement event
\ebcode{inc}. \ebcode{inc}'s guard states that the event is executed
when \ebcode{v} is to the right of the index \ebcode{r}, where \ebcode{f}
is a total function that is modelled as an array. The function
\texttt{prime} above is used in the definition of event
actions. Actions can either be deterministic or
non-deterministic. Deterministic actions use the symbol \ebcode{:=},
e.g., \ebcode{p := r+1} in Figure~\ref{fig_inc}. \eb provides two
non-deterministic action operators, namely, \ebcode{:$\,|$} (becomes
such that) and \ebcode{:$\in$} (becomes in). One could write the
deterministic action above as \ebcode{r:$\in$
  \lcb{}r+1\rcb}. Further, \ebcode{:$\in$} actions can always be
reduced to \ebcode{:$\,|$} actions. Hence, one could write the
previous non-deterministic action as \ebcode{r' = r+1}. Therefore,
\pyeb only supports non-deterministic actions built with the aid of
the \ebcode{:$\,|$} operator. We encode the right-hand side of our example
above as the \emph{before-after} predicate below.

\smallskip
\begin{lstlisting}[language=Python,morekeywords={assert,self}]
prime(self.p) == self.r+1
\end{lstlisting}
\smallskip

\noindent \textbf{Frame-conditions.} Likewise 
\emph{frame-conditions}~\cite{leino:dg:02} typically used in
behavioural interface specification languages~\cite{catano:chase:03}
in which one must state which part of the state a function or method
may modify, non-deterministic actions must account for the state
change of \emph{all} the machine variables. Therefore, we wrap up
before-after predicates within a \texttt{BAssignment} class which
includes two fields: the assigned machine variables and
the before-after predicate itself.

The whole event action is shown below, which corresponds to
\ebcode{inc}'s actions on the right side of
Figure~\ref{fig_inc}. \texttt{BAssignment}'s constructor takes the
before-after predicate as the second parameter and the set of machine
variables it modifies (assigns to) as the first parameter. Operator
\texttt{And} is Z3's logical-and operator.

\smallskip
\begin{lstlisting}[language=Python,morekeywords={assert,self}]
ba = BAssignment({self.p,self.q,self.r}, 
                 prime(self.p) == self.r+1)
\end{lstlisting}
\smallskip

The \texttt{skip} statement (see below) is implemented as a function
that takes a set \texttt{v} of machine variables and returns a
non-deterministic assignment such that no variable in \texttt{v} is
assigned to. The function \texttt{conjunct\_\-lst} returns the logical-and
of the predicates of a list.

  \smallskip
\begin{tabular}{c}
\begin{lstlisting}[language=Python,morekeywords={assert,self}]   
def skip(v):
 ba = conjunct_lst([(prime(elm) == elm) for elm in v])
 res = BAssignment(v,ba)
 return res
\end{lstlisting}
\end{tabular}
\smallskip

\noindent \textbf{Guards.} Event guards are modelled as Python
dictionaries. The event \ebcode{inc} (right side of Figure~\ref{fig_inc})
declares a single event guard, which is encoded as below, where
\ebcode{f} and \ebcode{v} are declared in the machine context, and
\ebcode{r} is a machine variable.


\smallskip
\begin{lstlisting}[language=Python,morekeywords={assert,self}]
guard = {'grd1': self.context.f(self.r) < self.context.v }
\end{lstlisting}
\smallskip

\noindent \textbf{Machine Events.} We use classes \texttt{BEvent} and
\texttt{BEventRef} to encode events and refinement events,
respectively. The status of an event is encoded as a Python enumerated
type (see below). Each event class declares a private class field of
type dictionary for storing the event guards. The dictionary follows
the structure of the object \texttt{guard} above, using strings as keys and
storing Z3 predicates. Each event class also has a class field of type
\texttt{BAssignment} for storing the event body and a field of type
\texttt{Status}. 

\smallskip
\begin{lstlisting}[language=Python]
Status = Enum('Status',
  ['Ordinary', 'Convergent', 'Anticipated']) 
\end{lstlisting}
  \smallskip

  \noindent \textbf{Machines.} Classes \texttt{B\-Machine} and
  \texttt{B\-Machine\-Refines} encode abstract and refinement machines, respectively. These classes include fields of type
  dictionary for modelling events, machine variables, and
  invariants. These classes also include a reference
  \texttt{\textbf{self}.context} to the machine
  context. \texttt{B\-Machine} and \texttt{B\-Machine\-Refines} use getters
  and setters to access and modify dictionaries.

\begin{figure}[t]
\begin{center}
\begin{tabular}{c@{\hspace*{12pt}}c}
  \begin{tabular}{l}
\ebkeyw{invariants}\\
~\ebtag{@inv1}~\ebcode{p $\in$ 1..n}\\
~\ebtag{@inv2}~\ebcode{q $\in$ 1..n}\\
~\ebtag{@inv3}~\ebcode{r $\in$ p..q}\\
~\ebtag{@inv4}~\ebcode{v $\in$ f[p..q]}\\
    ~\\
\ebkeyw{anticipated}\\
\ebkeyw{event}~\ebcode{progress}\\
\ebkeyw{then}\\
~\ebtag{@act1}~\ebcode{r :$\in{}\nat$}\\
\ebkeyw{end}
\end{tabular} &
\begin{tabular}{l}
\ebkeyw{convergent event}
\ebcode{inc}\\ 
\ebkeyw{refines}~\ebcode{progress}\\
\ebkeyw{where}\\
~\ebtag{@grd1}~\ebcode{f(r)$<$v}\\
\ebkeyw{then}\\
~\ebtag{@act1}~\ebcode{p := r+1}\\
~\ebtag{@act2}~\ebcode{r :$\in$r+1..q}\\
\ebkeyw{end}
\end{tabular} 
\end{tabular}
\end{center}
\caption{$i.)$ \ebcode{inc} abstract event~~~$ii.)$ \ebcode{inc} refinement event}
\label{fig_inc}
\end{figure}

\input{po}

%% file: po.tex
\subsection{Proof obligations}
\label{sec_po}

\pyeb generates proof obligations for contexts, machines, machine
refinements, events, and event refinements, following \eb's
semantics for proof-obligation
generation~\cite{abrial:pos:10}. Table~\ref{table:pos} summarises these proof obligations and describes the components
involved. The first column gives the name of the proof
obligation, the second column gives a small description, and the third
column shows the involved components.

\begin{table}[ht]
  \begin{centering}
    \begin{tabular}{|l|c|c|}
      \hline
      {\bf po}         & {\bf desc} & {\bf comp} \\ \hline\hline
      {\textsf{Th}}    & theorem    & context \rule{0pt}{2ex}\\ \hline
      {\textsf{Inv}}   & invariant preservation &  machine \rule{0pt}{2ex}\\ \hline
      {\textsf{Init}}  & initialisation         &  machine  \rule{0pt}{2ex}\\ \hline
      {\textsf{Fis}}   & feasibility            &  machine \rule{0pt}{2ex} \\ \hline
      {\textsf{Grd}}   & guard strengthening    &  machine refinement \rule{0pt}{2ex}\\ \hline
      {\textsf{Sim}}   & before-after predicate simulation             &  machine refinement  \rule{0pt}{2ex}\\ \hline
      {\textsf{Var}}   & variant decreasing     &  machine  \rule{0pt}{2ex} \\ \hline
      {\textsf{WFis}}  & witness feasibility   &  machine refinement  \rule{0pt}{2ex} \\ \hline
    \end{tabular} 
    \vspace{.1cm}
    \caption{Proof obligations}
    \label{table:pos}
  \end{centering}
\end{table}

The first row presents the \textsf{Th} proof obligation, which is generated in
the definition of a context, and which depends on the axioms and other theorems included
in that context. The invariant preservation proof obligation is defined for the
execution of an event, therefore, the event actions must set machine variables with
values that do not break the machine invariants. In particular, the
\ebkeyw{initialisation} event must provide machine variables with initial values for
which the machine invariants hold (\textsf{Init}).

The feasibility proof obligation (\textsf{Fis}) is defined for the event actions; therefore, every non-deterministic assignment must be feasible; in other words, there must exist a
valuation (a set of machine variables) that makes the before-after predicate used in the definition
of the action true.


Guard strengthening proof obligations (\textsf{Grd}) ensure that the
guard of a concrete event is stronger than the guard of an abstract
event. Simulation proof obligations (\textsf{Sim}) ensure that the
action of a concrete event simulates the action of the respective
abstract event; in other words, the before-after predicate of the
concrete event cannot contradict the before-after predicate of the
abstract event.

There is no requirement for \eb models to terminate and, in fact, most \eb models run
forever. However, we can use machine variants and convergent events to force
\eb models to terminate. An event status can be \ebkeyw{ordinary}, \ebkeyw{convergent},
or \ebkeyw{anticipated}. A machine \ebkeyw{variant} can be a natural number expression
or a finite set expression. A numeric \ebkeyw{variant} must be decreased by every
\ebkeyw{convergent} event, and a set \ebkeyw{variant} must be strictly included in
its previous value by all \ebkeyw{convergent} events. Hence, the variant decreasing
proof obligation (\textsf{Var}) ensures that every convergent event decreases the
machine variant. \ebkeyw{anticipated} events become \ebkeyw{convergent} in a machine
refinement. Hence, whereas for each \ebkeyw{convergent} event a proof obligation is
generated that states that the event decreases the machine \ebkeyw{variant}, for each
\ebkeyw{anticipated} event a similar proof obligation is generated, but for any
refinement event. \pyeb supports numerical variants, but does not support set
variants.


Refinement events can include \emph{witness} predicates for
\emph{disappearing} variables. A disappearing variable is a variable that
is in $x$ but not in $y$ in Figure~\ref{fig_evt}. For example, a refinement
(concrete) event can include a \ebkeyw{with}~$a:P(a)$ witness clause for a
disappearing variable $a$ and a predicate $P$. The witness feasibility proof
obligation (\textsf{WFis}) ensures that the predicate $P$ is feasible, that
is, there exist appropriate parameters that make $P$ true~\footnote{In
  general, $P$ depends on several parameters other than $a$ such as machine
  variables, sets, and constants.}.

In addition to the proof obligations in
Table~\ref{table:pos}, the Rodin platform generates well-definedness
proof obligations. These proof obligations intend to detect ill-formed
(theorems, axioms, functions, events, etc.)  definitions for which
unprovable proof obligations are generated. \pyeb does not generate
well-defined proof obligations; instead, it relies on the Z3's type
system to detect ill-formed formulae. Should \pyeb not rely on the Z3's type system, it would need to implement a bespoke syntax checker or parser.

%% file: exp.tex
\section{Case Study}
\label{sec_exp}
For our experimental evaluation of the \pyeb tool, we took various \eb
models for sequential algorithms, manually wrote Python programs for
them, and used the tool to verify them. The \eb models have been 
written by J.-R.~Abrial, and are available from
\href{https://web-archive.southampton.ac.uk/deploy-eprints.ecs.soton.ac.uk/122/} 
{https://\-web-\-ar\-chive.\-sout\-hamp\-ton.\-ac.\-uk/\-de\-ploy-\-e\-prints.\-ecs.\-so\-ton.\-ac.\-uk/\-122/}
as Rodin projects. They can be imported using the Rodin IDE~\cite{rodin:plat}. 

\pyeb takes a Python program as input and generates various types of
proof obligations as described in Section~\ref{sec_po}. \pyeb then
discharges each proof obligation or shows an error with the
information of the proof obligation that is unprovable. This error
information is a replica of the error information provided by the Z3
SMT solver. For our case study, Z3 was able to manage the generated
proof obligations by either issuing an error that we then used to evolve
the formal model or succeeding when the model was correct.

\medskip

\noindent
\textbf{Models of sequential programs}. In~\cite{abrialseq:09}, Jean-Raymond Abrial
describes an approach for designing and building sequential programs in \eb, therefore,
programs are represented as Hoare triples~\cite{Hoare69} composed of a program
precondition, the program itself, and a program postcondition. In this approach, the
program input is the set of constants defined in the machine context, the program
precondition is the set of axioms defined on those constants, the program postcondition
is the guard of a unique \emph{final} event defined in the last refinement machine, and
the program outputs are (the values of) the machine variables.

Software development with \eb is a three-stage process. Abstract events are written
during the \emph{specification} phase, event refinements are made during the \emph{design} phase,
and a final unique event (program) is calculated during the \emph{merging}
phase. The two former phases are strongly related to \pyeb. The latter
phase relates to the generation or synthesis of
code~\cite{catano:sttt:17}, which \pyeb does not support. Next, we
discuss the algorithms used in our case study as per the modelling
and verification of sequential programs. 

\medskip

\noindent
\textbf{Binary search}. According to its program precondition, the \eb model includes a
total function $f$ that models a vector of natural numbers, a positive constant $n$
modelling the size of $f$, and a value $v$ the algorithm searches for in $f$. The
vector $f$ is sorted. According to its post-condition, the algorithm finds the position
$r$ at which the value $v$ is. According to the algorithm design phase, the refinement
machines include indices $p$ and $q$, which hold the first and last positions of the subarray of $f$ over which the algorithm continues to search for $v$. The pivot
index $r$ is set halfway between $p$ and $q$. If $v$ is greater than $f(r)$, then
the algorithm searches $v$ to the right of $r$; otherwise, if the value is less
than $f(r)$, then the search continues to the left of $r$. Following Abrial's
approach to modelling sequential algorithms, the \eb model includes a \emph{final} event
whose guard is $f(r) = v$. 

\medskip

\noindent
\textbf{Minimum}. According to the program precondition, the \eb model includes a total
function $f$ that models a vector of natural numbers and a constant $n$
representing the number of elements of $f$. Regarding its post-condition, the algorithm
calculates the minimum element $m$ through an exhaustive search over vector $f$. According to the algorithm design phase, the refinement machines declare indices $p$
and $q$, whose initial values are $1$ and $n$, respectively. It declares two
events. An event searches from left to right, increasing the index $p$, and a second
event searches from right to left, decreasing the index. 


\medskip

\noindent
\textbf{Reversing}. As per its precondition, the \eb model declares a
total function $f$ whose size is a positive constant $n$. According to
its postcondition, the algorithm calculates a relation $g$ with
the same elements as $f$ but organised in reverse order.
According to the algorithm design phase, the model includes two
indexes, $i$ and $j$, with initial values $1$ and $n$, respectively. A
refinement event progresses by increasing $i$ and decreasing $j$. The
final event stops (its guard is) when $j\leq{}i$. The main reason we
did not include the reversing algorithm in our experimental evaluation
is that the event that progresses also uses some \emph{domain
  subtraction} operations (which are typical of \eb) that \pyeb does
not support.

\medskip


\noindent
\textbf{Search}. Regarding its precondition, the \eb model introduces a
function $f$ that models a vector of natural numbers, a positive constant $n$
modelling the size of $f$, and a value $v$ the algorithm searches for in $f$. The
algorithm does not require $f$ to be sorted. According to its post-condition,
the algorithm finds the vector index $r$ at which the value $v$ is. Regarding
the algorithm design phase, the sole machine refinement constrains $r$
to be between $1$ and $n$, and $f$ to contain $v$. The algorithm progresses
by giving $r$ an initial value of $1$ and increasing $r$ by $1$ each time $f(r)$ is different from $v$. The refinement machine includes a
\emph{final} event whose guard is $f(r) = v$, therefore, the final result is
$r$.



\medskip

\noindent
\textbf{Sorting}. According to its precondition, the \eb model declares a total
injective function $f$, and a constant $n$ modelling its size. According to its
post-condition, the algorithm calculates a relation $g$ whose domain is sorted and
has the same elements as $f$. The reason for not including the sorting algorithm in
our experimental evaluation is similar to that for not including the reversing algorithm
(discussed above).

\medskip

\noindent
\textbf{Square root}. According to its precondition, the model includes a non-negative
integer constant $n$ whose square root is to be calculated. As per its
postcondition, the algorithm finds a non-negative integer variable $r$ such that
$r\times{}r$ is less than or equal to $n$, yet $n$ is less than
$(r+1)\times{}(r+1)$. Therefore, $r$ is the integer square root of $n$. 


\medskip

\noindent
\textbf{Inverse}. As per its precondition, the model includes a
constant $n$, and a total function $f$ that is strictly increasing,
hence injective. $f$ is an integer-valued function. There exist two
values $a$ and $b$ such that $f(a)\leq{}f(n) < f(b+1)$. Regarding its
post-condition, the algorithm calculates a value $r$ such that $r$
is the inverse value of $n$ throughout the function $f$. Regarding the
algorithm design phase, machine refinement progresses by
increasing $r$ and decreasing $q$. The algorithm stops when $r$ and
$q$ are equal. 


Table~\ref{table:algos} shows some statistics related to the size and
structure of the algorithms included in the case study. The table has
five columns: the name of the algorithm, whether or not it has been
included in the case study, the number of \eb machines of the
algorithm, the type of algorithm and
the number of lines of code we have written for the respective Python
model.

\begin{table}[ht]
  \begin{centering}
    \begin{tabular}{|l|c|c|c|c|}
      \hline
      {\bf model} & {\bf \pyeb} & {\bf \# machine} & {\bf type} & {\bf \# loc}\\ \hline\hline
      {binary search}  & Yes & 3 & array & 180 \rule{0pt}{2ex}\\ \hline
      {minimum}     & Yes    & 2 & array & 115 \rule{0pt}{2ex}\\ \hline
      {reversing}   & No     & 2 & pointer & n/a  \rule{0pt}{2ex}\\ \hline
      {search}      & Yes    & 2 & array & 99  \rule{0pt}{2ex}\\ \hline
      {sorting}     & No     & 3 & array & n/a \rule{0pt}{2ex}\\ \hline
      {square root} & Yes    & 3 & numerical & 121 \rule{0pt}{2ex}\\ \hline
      {inverse}     & Yes    & 2 & array & 138 \rule{0pt}{2ex}\\ \hline
    \end{tabular} 
    \vspace{.1cm}
    \caption{Sequential models}
    \label{table:algos}
  \end{centering}
\end{table}

The Python versions of the algorithms are available from
\href{https://github.com/ncatanoc/pyeb/tree/main/sample}{https://\-git\-hub.\-com/\-n\-ca\-ta\-noc/\-pyeb/\-tree/\-ma\-in/\-sam\-ple}.

\input{binsearch}

\input{tool}

%% file: binsearch.tex
\subsection{The binary search algorithm}
\label{sec_binary}
Next, we discuss the object-oriented syntax and format that Python
classes should follow. A machine context should be declared as a
Python class and its constructor should initialise constants and
functions the usual way in Python, i.e., using the
\texttt{\textbf{self}} object syntax. Class constants and functions
should be encoded as Z3 constants and functions. Axioms and theorems
should be encoded as Python functions that return their predicate
encodings. These predicates are Z3 predicates. The names of all these
functions should be prefixed with \texttt{axiom\_} and
\texttt{theorem\_}, respectively. An abstract machine should be
declared as a Python class whose constructor is parametrised by a
context class object. A concrete machine should also be a Python
class, but its constructor is further parametrised by a reference to
the abstract machine. Events are declared within Python classes for
abstract and concrete machines. Events must be functions whose names
are prefixed with \texttt{event\_} and invariants with
\texttt{invariant\_}. Events in a refinement machine should be
prefixed with \texttt{ref\_\-event\_}.

In what follows, we present the Python encoding of the \eb excerpt in
Figure~\ref{fig_inc}, which is part of the binary search algorithm. The
figure omits the whole abstract and concrete \eb machines as well as their
contexts. Class \texttt{Context} includes the declaration of constants,
axioms, and theorems. Class \texttt{Context} declares class fields
\texttt{\textbf{self}.f}, \texttt{\textbf{self}.n}, and
\texttt{\textbf{self}.v} similar to the declarations at the beginning of
Section~\ref{sec_exp}.

Axioms are defined as functions that return their actual 
encoding. The Python encoding below shows the first axiom in
Figure~\ref{fig_inc}; therefore, every element in
\texttt{\textbf{self}.f} is non-negative. The axiom itself is
returned by the function \texttt{axiom\_\-axm1}. 

  \smallskip
\begin{tabular}{c}
\begin{lstlisting}[language=Python,morekeywords={assert,self}]
def axiom_axm1(self):
 x = Int('x')
 return (ForAll(x, Implies(And(x>=1, x<=self.n),
                           self.f(x)>=0)))
\end{lstlisting}
\end{tabular}
  \smallskip

Likewise axioms, theorems are coded as Python functions. They are defined in
class \texttt{Context}. Additionally, as explained in Section~\ref{sec_po},
theorems generate proof obligations that must be proven from the axioms
defined in such class.

  \smallskip
\begin{tabular}{c}
\begin{lstlisting}[language=Python,morekeywords={assert,self}]
def theorem_thm1(self):
 return (self.n>0)
\end{lstlisting}
\end{tabular}
  \smallskip

Class \texttt{Machine\_\-Binary\-Search\-\_ref0} is the abstract
machine. The class constructor sets initial values for the index
\texttt{\textbf{self}.r} and the machine context. Hence, the machine
context can be referred to from the machine.

  \smallskip
\begin{tabular}{c}
\begin{lstlisting}[language=Python,morekeywords={assert,self}]
def __init__(self,context):
 self.r = Int('r')
 self.context = context 
\end{lstlisting}
\end{tabular}
  \smallskip

  Events of the binary search abstract machine are introduced as
  functions. Event \texttt{event\_\-progress} is the most abstract event of
  the binary search algorithm. Its guard is the empty dictionary, which
  implies that the event is always enabled. Event \texttt{event\_\-progress}
  below non-deterministically assigns a non-negative value to the index
  \texttt{\textbf{self}.r}, mimicking the left side of
  Figure~\ref{fig_inc}. This event is refined by event \texttt{event\_\-inc}
  in class \texttt{Machine\_\-Binary\-Search\-\_ref1}. Event
  \texttt{event\_\-inc} is executed when value
  \texttt{\textbf{self}.\-context.\-v} is to the right of index
  \texttt{\textbf{self}.\-r}\footnote{\texttt{\textbf{self}.\-context.\-f} is
    ordered.}. Event \texttt{event\_progress} is \ebkeyw{anticipated}, it
  must thus be refined by a \ebkeyw{convergent} event in any refinement
  machine.

  \smallskip
\begin{tabular}{c}
\begin{lstlisting}[language=Python,morekeywords={assert,self}]
def event_progress(self):
 guard = {} 
 ba = BAssignment({self.r},prime(self.r) >= 0) 
 return BEvent('progress',Status.Anticipated,[],guard,ba) 
\end{lstlisting}
\end{tabular}
  \smallskip

  \texttt{Machine\_\-Binary\-Search\-\_ref1} is the first machine
  refinement. Its class constructor declares indices
  \texttt{\textbf{self}.p} and \texttt{\textbf{self}.q}, the leftmost
  and rightmost indices of vector
  \texttt{\textbf{self}.\-con\-text.\-f}. \texttt{Machine\_\-Binary\-Search\-\_ref1}'s
  class constructor sets the machine variant to
  \texttt{(\textbf{self}.\-q -\- \textbf{self}.\-p)}.

  \smallskip
\begin{tabular}{c}
\begin{lstlisting}[language=Python,morekeywords={assert,self}]
def __init__(self,abstract_machine,context):
 super().__init__(abstract_machine.context)
 self.context = context
 self.abstract_machine = abstract_machine
 self.p = Int('p')
 self.q = Int('q')
 self.variant = (self.q - self.p) 
\end{lstlisting}
\end{tabular}
\smallskip



  

  Event \texttt{ref\_\-event\_\-inc} refines abstract event
  \texttt{ref\_\-event\_\-progress}. The abstract event is
  \ebkeyw{anticipated} so the refinement event is
  \ebkeyw{convergent}. Event \texttt{ref\_\-event\_\-inc} must
  comply with the \ebkeyw{variant} of the refinement
  machine. Event \texttt{ref\_\-event\_\-inc}'s guard states that
  the event is executed when the value
  \texttt{\textbf{self}.\-con\-text.\-v} is at the right of index
  \texttt{\textbf{self}.\-r}, in which case
  \texttt{\textbf{self}.\-p} is set to
  \texttt{\textbf{self}.\-r+1}, \texttt{\textbf{self}.\-r} takes a
  non-deterministic value between \texttt{\textbf{self}.\-r+1} and
  \texttt{\textbf{self}.\-q}. The expression
  \texttt{prime(\textbf{self}.q) == \textbf{self}.q} states that
  \texttt{\textbf{self}.q} remains unchanged. Notice that this is
  not explicitly stated on the right side of Figure~\ref{fig_inc}.

  \smallskip
\begin{tabular}{c}
  \begin{lstlisting}[language=Python,morekeywords={assert,self}]    
def ref_event_inc(self):
 guard = {'grd1': (self.context.f(self.r) < self.context.v)}
 ba = BAssignment({self.p, self.q, self.r},
     And(prime(self.p) == self.r+1,
         prime(self.r) >= (self.r+1), prime(self.r) <= self.q, 
         prime(self.q) == self.q))
 inc = BEventRef('inc', super().event_progress())
 inc.set_status(Status.Convergent)
 inc.add_guards(guard)
 inc.add_bassg(ba)
 return inc
\end{lstlisting}
\end{tabular}
\smallskip

\eb machines have a distinguishable event, the
\ebkeyw{initialisation} event, which initialises machine
variables. As per the first refinement machine, the initial value
of \texttt{\textbf{self}.\-p} is \texttt{1}, the initial value of
\texttt{\textbf{self}.\-q} is the size of vector
\texttt{\textbf{self}\-.con\-text.\-f}, and index
\texttt{\textbf{self}.\-r} takes a non-deterministic initial value
between these two values. The event guard is the empty dictionary;
thus, the event is always enabled.

  \smallskip
\begin{tabular}{c}
\begin{lstlisting}[language=Python,morekeywords={assert,self}]        
def ref_event_initialisation(self):
 guard = {} 
 ba = BAssignment({self.p,self.q,self.r},
     And(prime(self.p) == 1, prime(self.q) == self.context.n,
         prime(self.r) >= 1, prime(self.r) <= self.context.n))
 init = BEventRef('initialisation', 
                  super().event_initialisation())
 init.set_status(Status.Ordinary)
 init.add_guards(guard)
 init.add_bassg(ba)
 return init  
\end{lstlisting}
\end{tabular}
  \smallskip

%% file: tool.tex
\subsection{Tool Usage}
\label{sec_tool}

\pypi~\cite{pypi} is a repository for the Python language which users
can use to publish Python libraries, made available to the entire
community of Python users. \pyeb is a \pypi library
implementation. \pyeb is installed with \pip~\cite{pip} via the
\texttt{python3\- -m\- pip\- install\- pyeb} command line. This line
further installs the \texttt{z3-solver},
\texttt{antlr4-python3-runtime} and \texttt{antlr4-tools}, which are
required packages. \pyeb requires the first package to discharge the
proof obligations automatically (see Section~\ref{sec_po}), and the
remaining two packages to translate Python programs into a sequence of
interactions the Z3 SMT solver then discharges.

\pyeb's source code is hosted at GitHub, reachable from
\href{https://github.com/ncatanoc/pyeb}
{https://\-github.\-com/\-n\-catanoc/\-pyeb}; the GitHub site includes
a \texttt{sample} folder with the Python algorithm files of the
experimental evaluation presented in Section~\ref{sec_exp}. We
manually wrote these Python algorithms directly from the respective
\eb versions of J.-R. Abrial in~\cite{abrialseq:09}.

For example, you can run the Python model of the binary search algorithm by
either typing \texttt{pyeb \-sample/\-bin\-search\_\-oo.py} or by executing
\pyeb as a module or package via \texttt{python3 -m pyeb
  sample/\-bin\-search\_\-oo.py}. Either command line creates a
\texttt{bin\-search\_\-oo\_\-obj.\-py} object Python file that includes a
sequence of object creation instructions and calls to the Z3 SMT
solver.

%% file: related.tex
\section{Related Work}
\label{sec_rel}
The authors in~\cite{rocha:eb2py:20} propose translating \eb into
the Python programming language. The translation is implemented as a
Rodin plugin that generates a design-by-contract Python program that
uses \texttt{assert} and \texttt{raise} Python statements to verify
the contracts at runtime. Their work relates to ours, but we will
instead generate code \emph{outside} the Rodin IDE.

The authors in~\cite{cai:ml:repair:19} propose a joint machine learning
and \eb methodology to automatically \emph{repair} faulty \eb
models, for example, models that include deadlocked states or events
that break machine invariants. Although our work with \pyeb strives to check
the soundness of a model, a complementary and related aspect is to
repair unsound models, for which machine learning and
program repair techniques seem promising to us.

\mypy~\cite{mypy} is a static type checker for Python that checks if
program variables and functions are used with the right type. It
issues a warning in case they are not. \pylint~\cite{pylint} is a
static Python tool that looks for code smells and makes suggestions on
code refactoring and good programming
practices. \pyflakes~\cite{pyflakes} is a light version of \pylint; it
is simpler, faster, and does not compile the source program. These 3
Python static checkers are \pypi library
implementations~\cite{pypi}. Our \pyeb tool is not a Python analyser,
but rather allows programmers to write mathematical models in Python
of programs that can eventually be implemented in Python or any other
programming language. \pyeb could use type annotations and
compile-time type checking to assist a (hypothetical) code generation
process, but this is still future work.

Other static analysers for Python are
\prospector~\cite{prospector}, built on top of \pylint, and
\pyflakes, which can suppress spurious warnings; also, the
\bandit tool~\cite{bandit}, a tool designed to find common security
issues in Python programs from their AST (Abstract Syntax Trees)
representations.

The \nagini~\cite{nagini:18} tool is an automatic verifier for statically typed
Python programs. \nagini builds on top of \viper. \nagini\ verifies
memory safety, termination, absence of deadlocks, and functional
properties of Python programs. \nagini\ translates a Python program
and its specifications into the \viper\ language~\cite{viper:17}, for
which automated verifiers already exist. \nagini\ requires input
programs to adhere to the PEP 484 standard syntax for type
annotations, which also the \mypy\ tool implements.

\pymc~\cite{pymc,pymc:21} is a model checker of Python programs. It is also
implemented as a \pypi library. It takes a program written in
Python and transforms it into a Kripke structure, which is then
model-checked with Maude~\cite{maude:03}. \pymc\ supports LTL and CTL
temporal logics.

%% file: conclusion.tex
\section{Conclusion and Future Work}
\label{sec_conc}

This paper presented \pyeb, a refinement-calculus Python
implementation. \pyeb is implemented as a \pip
library. \pyeb supports \eb's syntax, including the
definition and use of events, contexts, machines, machine refinements,
machine variants and invariants, and witness clauses. Variants are
used to enforce termination. We carried out a case study in which
various sequential algorithms are modelled in Python from existing \eb
versions. \pyeb was able to deal with special \eb constructs such as
non-deterministic assignments, and machine variants,
which are at the core of formal software development with \eb.

Our work with \pyeb is part of a long-term work in which we plan to
implement a
\emph{Correct-\-by-\-Construction}~\cite{watson:cbc:16}
(CbC) framework to generate Python and Rust \emph{certified} code from
\eb models written as Python programs. The Python programs shall 
follow the syntax described in this paper. This CbC work will be based
on previous work on program synthesis for \eb and the Java programming
language presented in~\cite{catano:sttt:17,catano:EB2Java:14}. As \eb
models are models of reactive systems, challenges for this programme
synthesis work would be related to the code generation of high-performance implementations that are correctly synchronised.

As Z3 does not provide native support for all the plethora of sets and
relation operators the \eb language ships, we will investigate ways
\pyeb can fully provide support for those operators. As an
alternative, in previous work~\cite{catano:yices:11}, we implemented a
Yices library that included most of \eb's set and relational
operators, including domain subtraction, domain restriction, function
overriding, etc. We plan to either hook \pyeb into the existing Yices
library implementation or port the library directly to Z3's
language.

Functional correctness and software safety are key to software
development. A software system is safe if it does not exhibit some bad
behaviour. The safety properties in \eb are modelled as machine
invariants. We plan to extend \pyeb supported syntax for users to
write and verify LTL~\cite{Pnueli1977} (Linear Temporal Logic)
invariant properties in Python. Temporal logic allows one to model
safety (``something bad does not occur'') or liveness (``something
good will eventually happen'') security requirements.  Verifying
liveness properties in the context of a design-by-contract
specification language has already been studied
elsewhere~\cite{groslambert:liveness:JML:08}.